\documentclass[prb,aps,twocolumn, superscriptaddress]{revtex4}
\usepackage{graphicx, amsmath, amsthm, subfigure}
\makeatletter
\begin{document}
\title{Current induced local spin polarization due to the spin-orbit coupling in a two dimensional narrow strip}
\author{Qian Wang, L. Sheng, and C. S. Ting}
\affiliation{Texas Center for Superconductivity, Univerity of Houston, Houston, Texas 77204}
\date{today}
\begin{abstract}
The current induced local spin polarization due to weak Rashba spin-orbit coupling in 
narrow strip is studied. In the presence of longitudinal charge current, local spin polarizations 
appear in the sample. The spin polarization perpendicular to the plane  has opposite 
sign near the two edges. The in-plane spin polarization in the direction 
perpendicular to the sample edges also appears, but does not change sign across the 
sample.  From our scaling analysis based on increasing the strip width, the out-of-plane spin polarization is
important mainly in a system of mesoscopic size, and thus appears not to be associated with the spin-Hall effect in 
bulk samples. 
\end{abstract}
\maketitle

In a spin-orbit coupled electron system, an external electric
field can induce a transverse spin current, giving rise to the
so-called spin Hall effect (SHE). The SHE may offer a new way to
control electron spins in semiconductors, and so have potential
applications in spintronic devices. Depending on its origin, the
SHE is generally divided into two categories: the extrinsic SHE,
which originates from spin-dependent electron anomalous scattering
by impurities, and the intrinsic SHE, which occurs even in the
absence of impurities. The extrinsic SHE, was first proposed by
D'yakonov and V. I. Perel~\cite{yakonov71} in 1971 and reexamined
recently by Hirsch~\cite{Hirsch99} and Zhang~\cite{Zhang00}. The
intrinsic SHE was predicted by Murakami, Nagaosa, and
Zhang~\cite{Zhang03} for p-type semiconductors and by Sinova et
al.~\cite{Sinova04} for n-type semiconductors in two-dimensional
heterostructures. The intrinsic SHE has attracted much theoretical
interests~\cite{t3,t4,t5,t6,t7,t8,t9,t10,t11,t12,t13,t14,t15}.
Very recently, two independent groups have reported experimental
evidence~\cite{Kato,Wunderlich} that an electric field can cause
out-of-plane spin accumulations of opposite sign on opposite edges
of semiconductor films, which is considered to be a signature of
the SHE. Several analytical and numerical works have been published on the
subject of spin accumulation in a semiconductor with spin-orbit
coupling. Governale and Z\"ulicke~\cite{Zulicke02} were the first
to investigate spin accumulation. They studied the spin structure
of electron states in a quantum wire with parabolic confining
potential and strong Rashba spin-orbit coupling. Usaj and
Balseiro~\cite{Usai} showed that in a semi-infinite system with
spin-orbit coupling, a current flowing parallel to the edge
induces a net magnetization close to the edge. Using the Landauer-B\"{u}ttik formula for 
a tight-binding model, Nikoli\'{c} {\it et al.}~\cite{Nikolic} showed numerically that in a two-dimensional
bar with a width of 30 lattice constant, the Rashba spin-orbit coupling can induce opposite spin
accumulation near the two edges, which is qualitatively similar
to that observed in the experiment. In order to clarify whether
such spin polarizations are related to the SHE, it is important to
investigate their scaling behavior with increasing sample
size, and to
reveal the parameters that control the relative magnitude of the
 spin accumulation or polarization. 

In this paper, the electron wave function in a continuous
model is obtained for an infinite long conducting strip with finite width $L$.
Using the Kubo formula, we show that a longitudinal electrical
current induces both out-of-plane spin polarization ($S_z$) and
in-plane spin polarization ($S_y$). Near the two edges, the spin
polarization $S_z$ has opposite sign, whereas $S_y$ has the same
sign. When sample width $L$ increases, its scaling behavior indicates that $S_z$ near the edges
decreases and $S_y$ becomes dominant for given fixed electrical
current density. Therefore, the out-of-plane spin polarization is
an effect due to boundary reflections from the two opposite edges, and appears not to be
related to the SHE in a bulk sample.

Let us consider a system of a two-dimensional (2-D) infinite
long conducting strip with finite width $L$. The Hamiltonian for
the system with Rashba spin-orbit coupling can be written as  by
\begin{equation}\label{hamil}
H=\frac{k^2}{2m}+\lambda(\sigma_xk_y-\sigma_yk_x),
\end{equation}
where $\lambda$ is the coupling constant of  spin-orbit interaction, $\sigma_x$ and $\sigma_y$ are the Pauli matrices, $m$ is the electron effective mass, and we take units with $\hbar=1$.

The eigenstates of plane waves are
\begin{equation}\label{hstate}
|E_\mp,\vec k>=|E_\mp,k_x,k_y>=\frac{1}{\sqrt{2}}e^{i{\vec k}\cdot{\vec r}}{\pm ie^{-i\phi}\choose 1},
\end{equation}
where $\phi=\arctan(k_y/k_x)$, $+$ ($-$) labels lower (higher) energy eigenstate  with eigenvalue $E_\pm=k^2/2m\mp\lambda k$ for a given $\vec k$.

Assuming hard-wall boundary conditions, the wave function at the two edges ($y=0$ and $y=L$) is zero. Since the system is uniform along the $x$ direction, $k_x$ commutes with the Hamiltonian and is a good quantum number. We can write a eigenstate, with eigen-energy $E$, of the system as a superposition of four plane waves, with same $E$ and $k_x$. Suppose the system is in universal region as defined in Ref.~[5], the wave function near the Fermi level is given by
\begin{equation}\label{wf}
\begin{split}
\Psi(E,k_x,y)&=|E,k_x>\\&=\alpha_{k_x}|E,k_x,k_{y}^->+\beta_{k_x}|E,k_x,-k_{y}^->\\
&+\gamma_{k_x}|E,k_x,k_{y}^+>+\delta_{k_x}|E,k_x,-k_{y}^+>,
\end{split}
\end{equation}
where $k_{y}^\pm=\sqrt{k^{\pm2}-k_x^2}$ and $k^\pm=\pm\lambda m+\sqrt{\lambda^2m^2+2m E}$
with boundary conditions
$\Psi(k_x,0)=\Psi(k_x,L)=0.$
One can solve the boundary conditions and find the eigenvalues of $k_x$, which are a discrete set of values in the 
interval of $(-k_F^+,k_F^+)$, here $k_F^+=k^+$ with $E=E_F$ (the Fermi energy).	 In Usaj and Balseiro's work, there is only one edge, the eigenfunctions 
are propagating waves written as a superposition of one incident and two reflected waves. $k_x$ can take any value between $(-k_F^+,k_F^+)$. In our current study, the interference due to the two edges of the strip limits number 
of eigenvalues for $k_x$ at Fermi level, which  could inject rather different physics for the problem. 

While the four plane waves have different spin polarizations within the
two-dimensional plane, their interference leads to nonzero
out-of-plane local spin density. The local spin polarization
depends on the sign of the conserved longitudinal wave vector
$k_x$. For any given energy $E$ and a positive eigenvalue $k_{Enx}$
for $k_x$, $-k_{Enx}$ is also an eigenvalue for $k_x$.  In the
ground state where both positive and negative $k_x$ states are
occupied, the total local spin density is zero since the
contribution of each spin band is zero. However, if there are a
longitudinal current flowing in the strip, which causes a small
shift of the Fermi circles. The numbers of occupied states with
positive $k_x$ and negative $k_x$ are no longer equal, which can
induce net spin polarizations in the strip.

The net local spin polarization can be calculated
using Kubo formula~\cite{Kubo,Crepieus01},
\begin{equation}
\begin{split}
&\frac{\vec S(y)}{\cal{E}}=\frac{ie}{V}\sum_{k_x,E,E^\prime}(f_{E^\prime}-f_{E})\\
&\times\frac{[<E^\prime,k_x|\frac{1}{2}\vec\sigma(y)|E,k_x> <E,k_x|v_x|E^\prime,k_x>]}{(E-E^\prime
)(E-E^\prime-i\delta)}.
\end{split}\label{SPKUBO}
\end{equation}
where $\cal{E}$ is the electric field, and $v_x$ is given by
$v_x=k_x/m-\lambda \sigma_y$. Here, we wish to point
out that the contribution to the spin polarization comes from the Fermi
surface, in contrast to the intrinsic SHE, which comes from
the contribution of all occupied states. It is easy to find that
the spin polarization given by Eq\ (\ref{SPKUBO}) diverges in the
clean limit. To overcome this problem, we consider a constant
longtidudinal current density $I/L$ is driven through strip. We
calculate the ratio between the spin polarization and the current
density
\begin{widetext}
\begin{equation}
\frac{\vec S(y)L}{I}=
\frac{1}{e}\frac{\sum_{k_x,E,E^\prime}[<E^\prime,
k_x|\frac{1}{2}\vec\sigma(y)|E,k_x> <E,
k_x|v_x|E^\prime,k_x>]\delta(E^\prime-E)\delta(E-E_F)}
{\sum_{k_x,E,E^\prime}[<E^\prime, k_x|v_x|E, k_x><E,
k_x|v_x|E^\prime,k_x>]
\delta(E^\prime-E)\delta(E-E_F)}\ ,
\end{equation}
\end{widetext}
which is a finite quantity.

All the coefficients in Eq.~\ref{wf} can be determined
numerically. However, we found that the eigenfunctions (standing
waves) can not be expressed as a superposition of the two
eigenfunctions~\cite{Usai} of same $k_x$ obtained in the case of
only one edge. We also found that, for $|k_x|<k_F^-$,
$|\alpha|=|\beta|$ and $|\gamma|=|\delta|$; for
$k_F+>|k_x|>k_F^-$, $|\gamma|=|\delta|$. 

\begin{figure}[tbp]
    \centering
    \subfigure[]{
        \includegraphics[height=1in,width=3in]{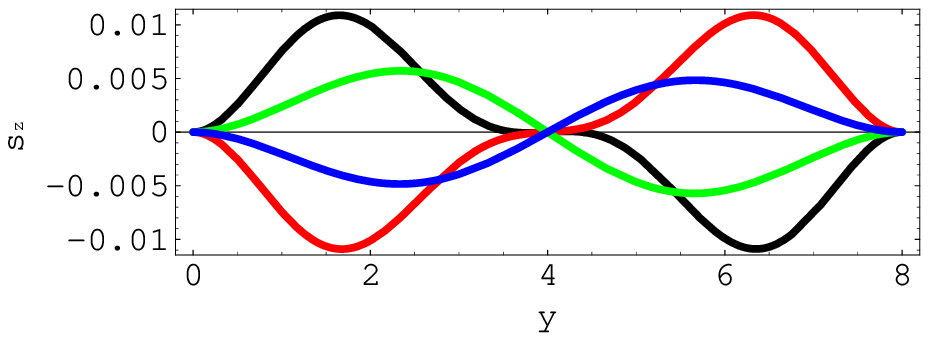}
    \vspace{0.03in}}
    \subfigure[]{
    \includegraphics[height=1in,width=3in]{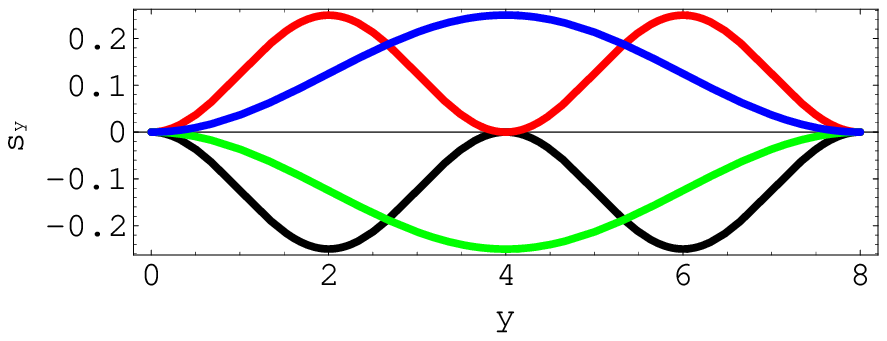}
    \vspace{0.03in}}
    \subfigure[]{
        \includegraphics[height=1in,width=3in]{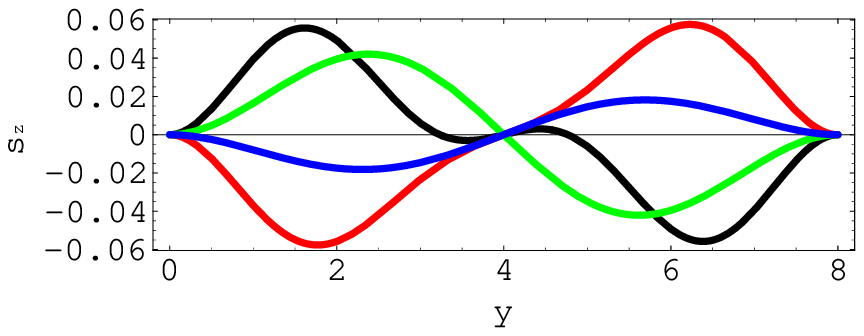}
    \vspace{0.03in}}
    \subfigure[]{
        \includegraphics[height=1in,width=3in]{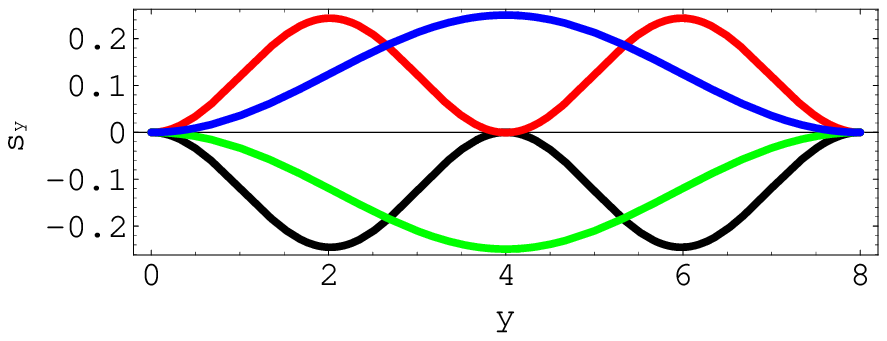}
    \vspace{0.03in}} 
    \subfigure[]{
        \includegraphics[height=1in,width=3in]{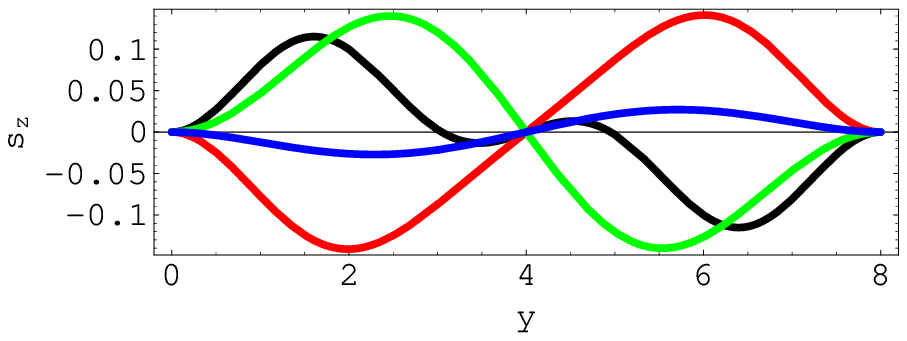}
        \vspace{0.03in}}
    \subfigure[]{
        \includegraphics[height=0.9in,width=3in]{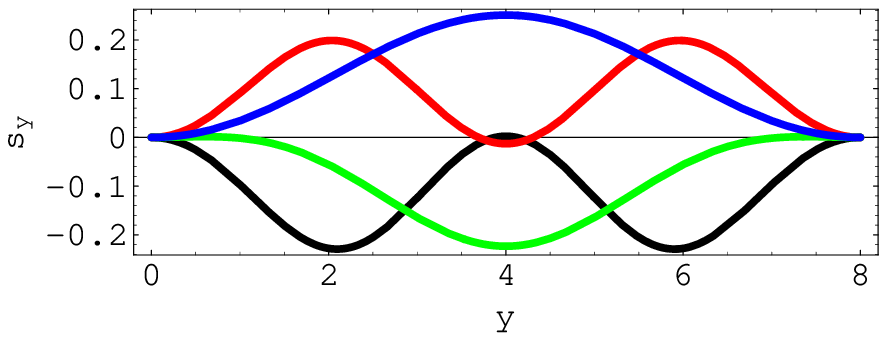}}
    \caption{$s_z$ [figure (a), (c), and (e)] and $s_y$ [figure (b), (d), (f)] as
    function of position $y$ for eigenvalues of positive $k_x$ at the Fermi level.
    $s_z$ and $s_y$ are in units of $\frac{\hbar}{2}$.
    $L=8/k_F$, where $k_F$ is the Fermi wave vector when there is no spin-orbit
    coupling. For each of the three values of $\lambda$, there are 4 eigenvalues of $k_x$.
    In (a) and (b), $\lambda=0.01$, $k_x=0.609$ (black line), 0.628982 (red line), 0.909672
    (green line), and 0.929663 (blue line) ; in (c)(d), $\lambda=0.05$, $k_x$=0.570027
    (black line), 0.667649 (red line), 0.870449 (green line), and 0.969279 (blue line);
    in (e) and (f), $\lambda=0.1$, $k_x=0.526814$ (black line), 0.703971 (red line), 0.830598
    (green line), and 1.0172 (blue line)}.
    \label{fig:wf}
\end{figure}

Plots in Fig.~\ref{fig:wf} show local spin polarizations $s_z=\frac{\hbar}{2}<E_F,k_{x}|\sigma_z|E_F,k_{x}>$ and $s_y=\frac{\hbar}{2}<E_F,k_{x}|\sigma_y|E_F,k_{x}>$ as
functions of position $y$ in all the eigenstates that have
positive $k_x$ with eigen-energy at the Fermi level for three
different values of Rashba coupling $\lambda$, which is in the units of $k_F$. The width of the
strip is set to be $8/k_F$, where $k_F$ is the Fermi wave vector
when there is no spin-orbit coupling. $k_F$ is related to electron
density in the sample by $k_F^2=2\pi n$. Using typical value of
two-dimensional electron density $10^{12}$
cm$^{-2}$ [see Ref. 26], we estimate $k_F\approx 10^8/m$ and L is
around 80 nm. The spin polarizations $s_z$ and $s_y$ vanish at the
two edges as required by the boundary conditions. For
each eigenstate, $s_z(y)=-s_z(L-y)$ , whereas $s_y(y)=s_y(L-y)$. We
have also obtained $s_x$ and it is zero across the sample. Without
spin-orbit coupling, $k_y$ is quantized to values $k_{yn}=n\pi/L$,
where $n=1,2,3,\cdots$. For each $k_{yn}$, the eigenstates for two
spin directions are degenerate. When spin-orbit coupling is in presence, the two spin bands are no longer degenerate. However, the spin polarization increases as the Rashba
coupling increases. When we further increase $\lambda$, some
values of $k_x$ are larger than Fermi wave vector of the higher
spin band, as shown in Figs.~\ref{fig:wf}(e) and (f). Under this case, 
decaying waves show up in the wave functions along the y-direction for the higher spin band.

\begin{figure}
    \centering
    \subfigure[]{
        \includegraphics[height=0.9in,width=1.5in]{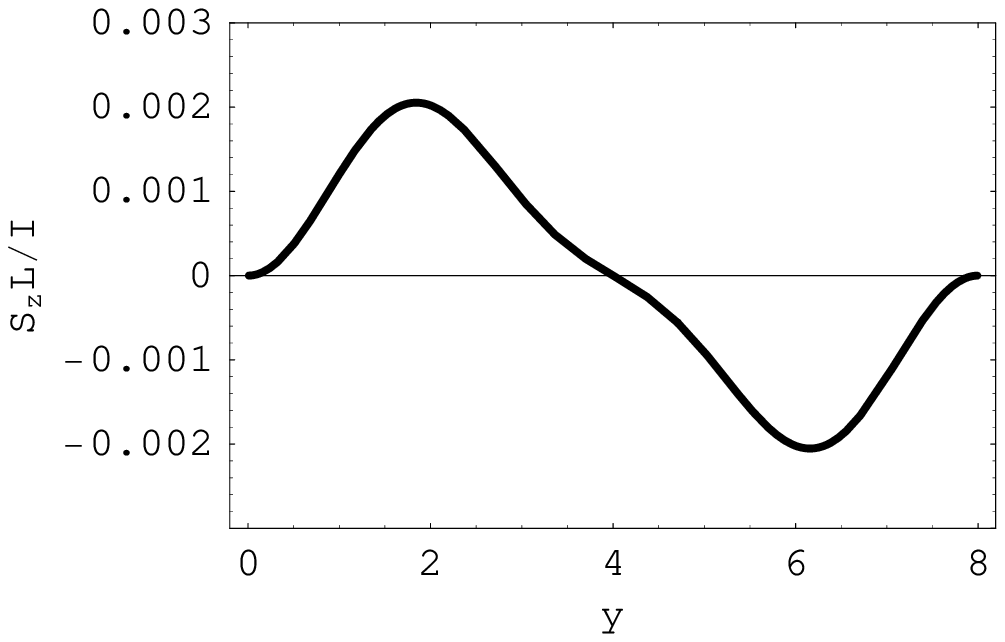}}
    \subfigure[]{
    \includegraphics[height=0.9in,width=1.5in]{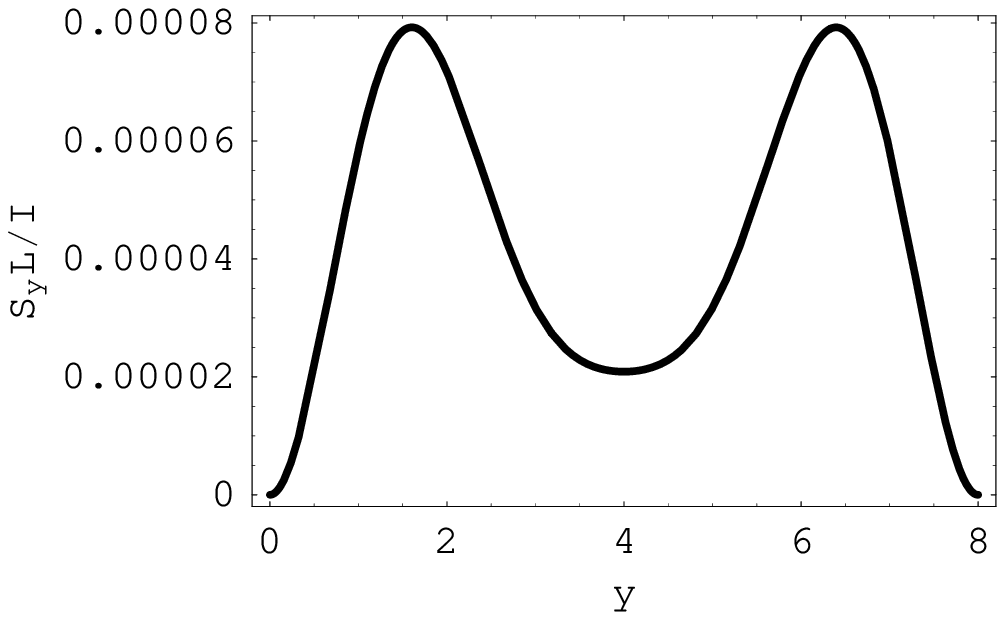}}
        \vspace{0.03in}
    \subfigure[]{
        \includegraphics[height=0.9in,width=1.5in]{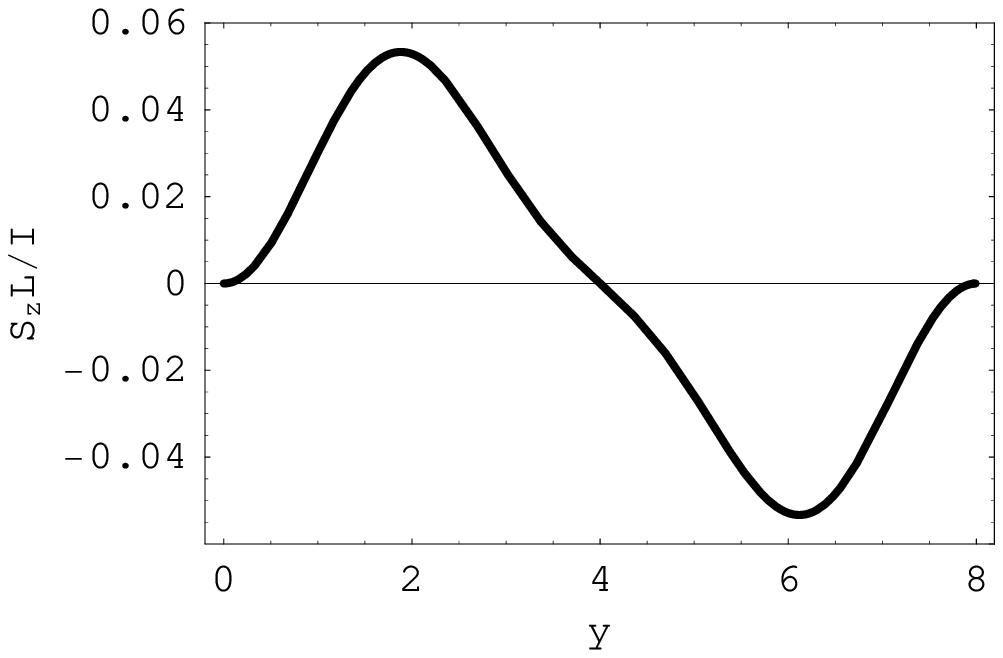}}
        \subfigure[]{
        \includegraphics[height=0.9in,width=1.5in]{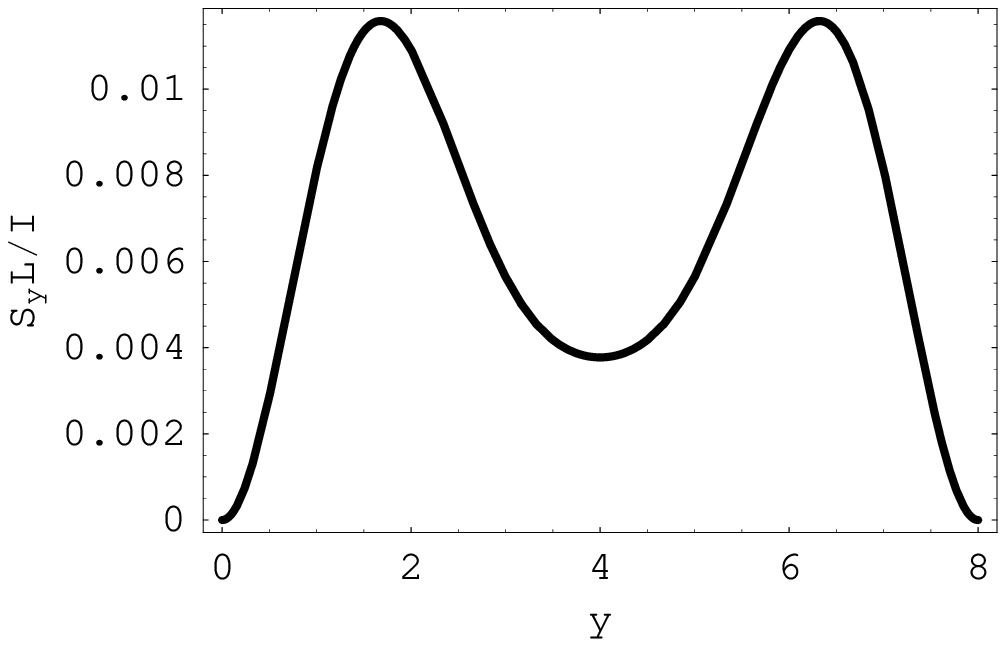}}
        \vspace{0.03in}
    \subfigure[]{
        \includegraphics[height=0.9in,width=1.5in]{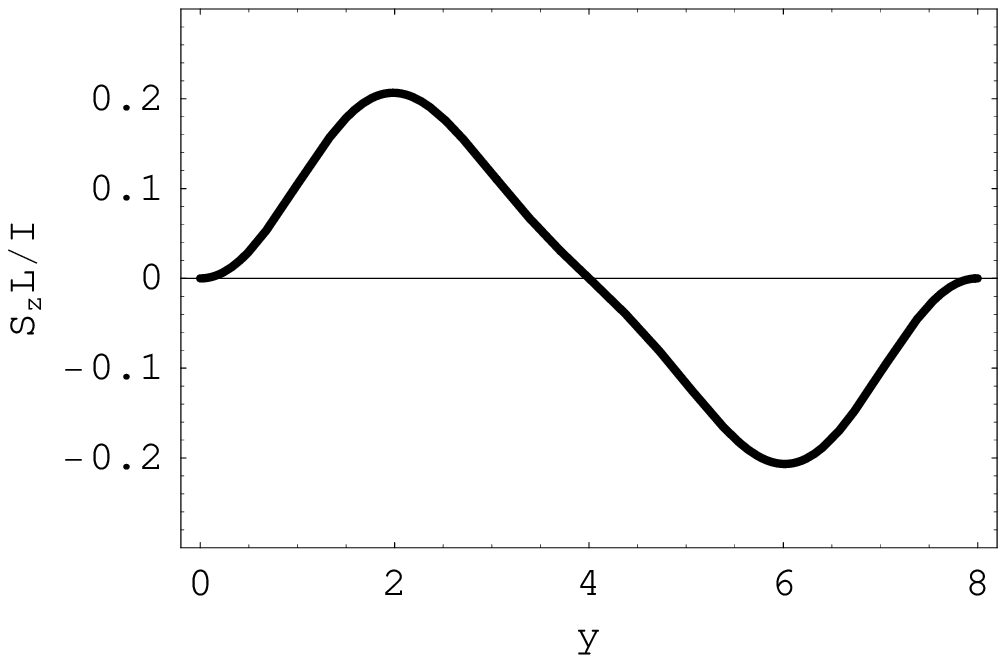}}
        \subfigure[]{
        \includegraphics[height=0.9in,width=1.5in]{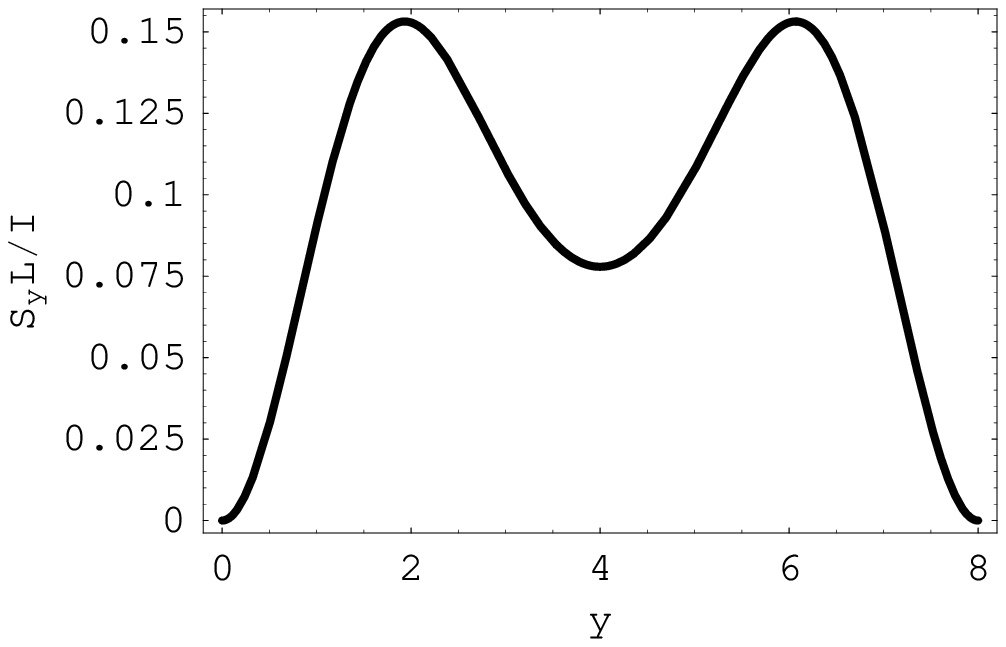}}
        \vspace{0.03in}
    \caption{Total local $S_zL/I$ and $S_yL/I$ as a function of y for $L=8/k_F$. In this and following figures they are in units of $\hbar/(2ek_F^2)$
    In (a) and (b) $\lambda=0.01$. In (c) and (d), $\lambda=0.05$. In (e) and (f), $\lambda=0.1$. }
    \label{fig:totalL8}
\end{figure}

The net spin polarizations are calculated by using the Kubo
formula in Eq.\ (\ref{SPKUBO}). Figures~\ref{fig:totalL8}(a), (c) and
(e) show the net $S_z(y)L/I$ when we sum the contribution from all
the positive $k_x$ modes at the Fermi level. The longitudinal
charge current also induces a local in-plane polarization
$S_y$~\cite{t5,Nikolic}, as shown in Figs.~\ref{fig:totalL8}(b),
(d) and (f), whereas $S_x\equiv 0$. Unlike $S_z$, $S_y$ has the
same sign across the sample. At weak Rashba couplings
$\lambda=0.01$ or $0.05$, we see from Figs.~\ref{fig:totalL8}
(a)-(d) that $S_z$ is one or two order greater than $S_y$ in
magnitude. With increasing the Rashba coupling, $S_y$ increases
much faster than $S_z$. As a consequence, $S_y$ becomes comparable
to $S_z$ at relatively large Rashba coupling $\lambda=0.1$, as
seen from Figs.~\ref{fig:totalL8}(e) and (f).
\begin{figure}[tbp]
    \centering
    \vspace{0.03in}
    \subfigure[]{
        \includegraphics[height=0.9in,width=1.5in]{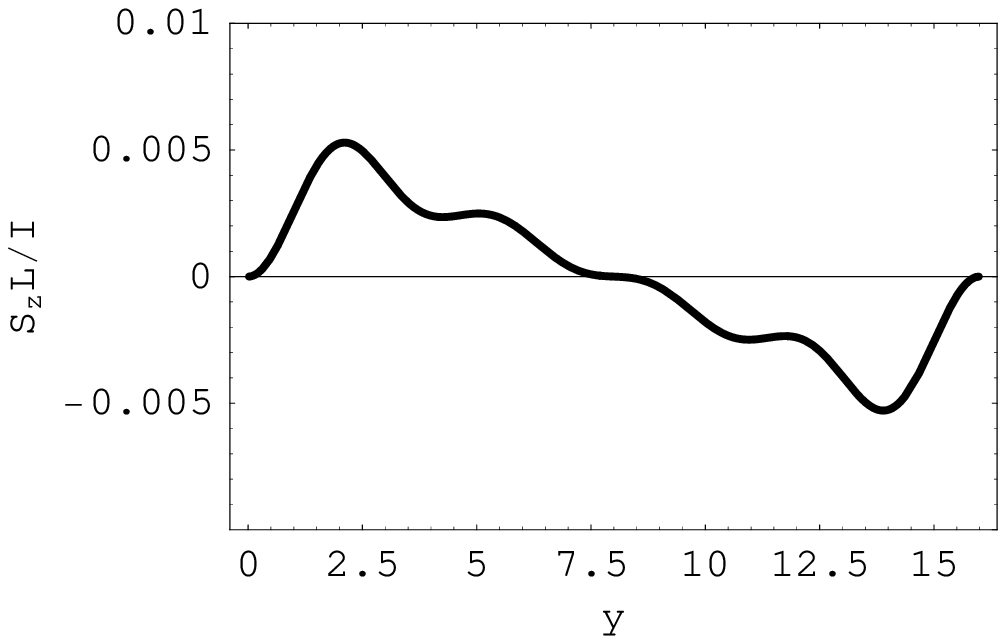}}
        \subfigure[]{
        \includegraphics[height=0.9in,width=1.5in]{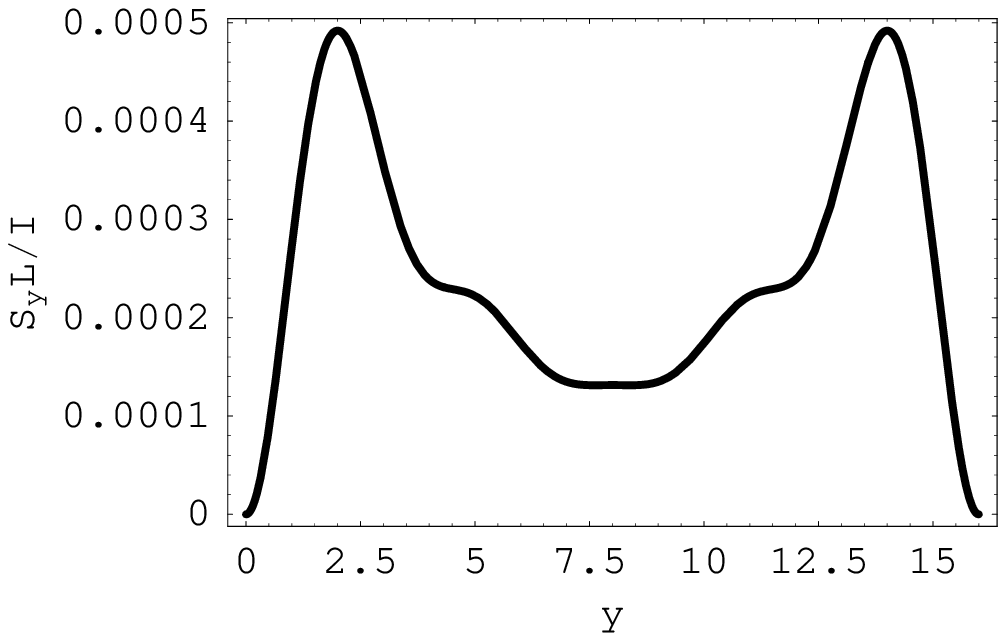}}
        \vspace{0.03in}
    \subfigure[]{
        \includegraphics[height=0.9in,width=1.5in]{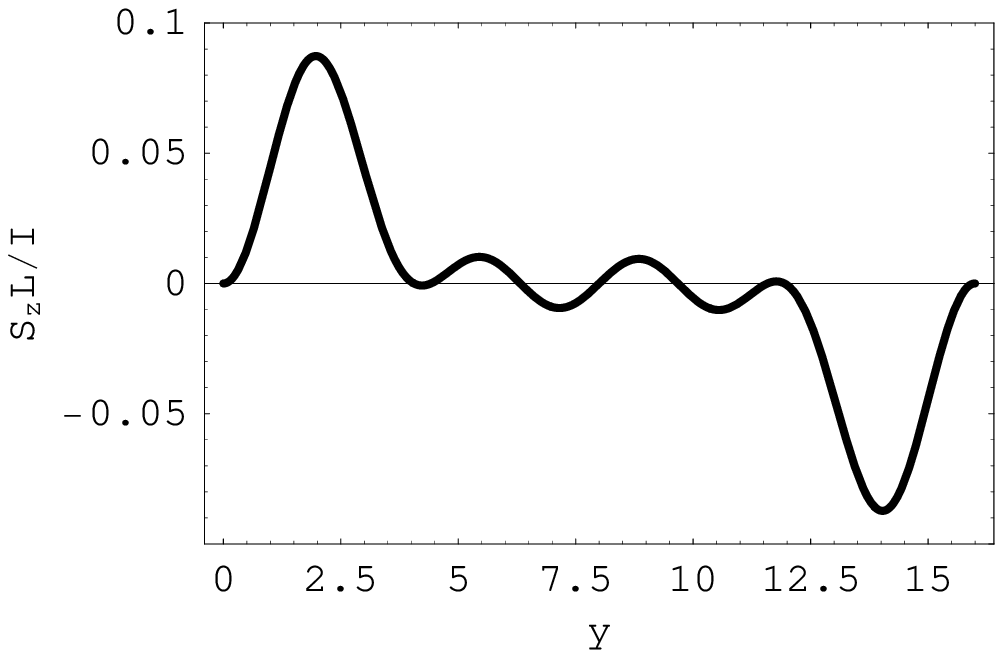}}
        \subfigure[]{
        \includegraphics[height=0.9in,width=1.5in]{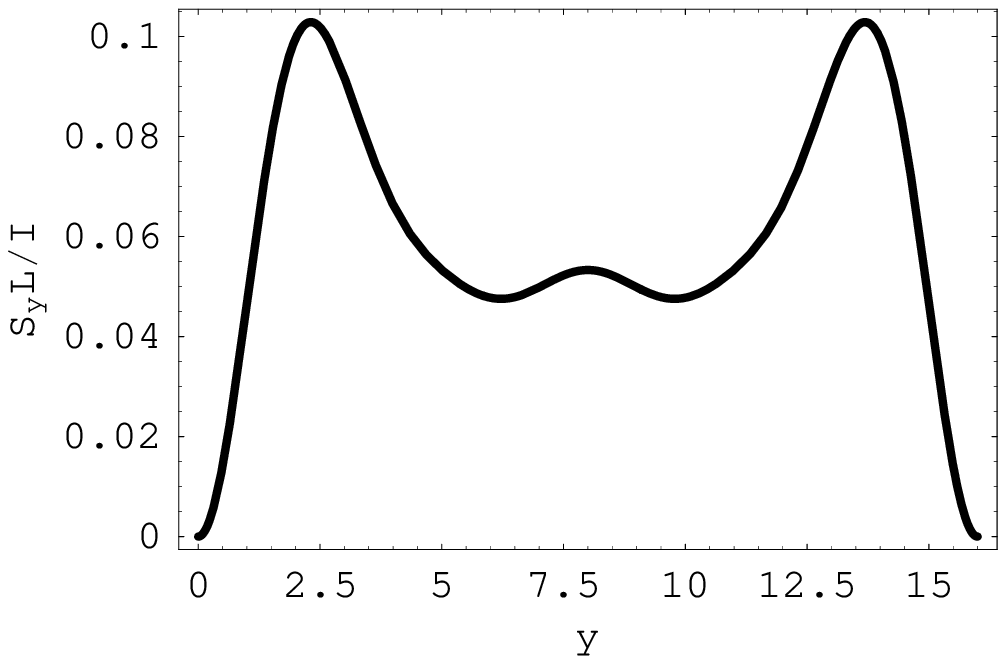}}
        \vspace{0.03in}
    \subfigure[]{
        \includegraphics[height=0.9in,width=1.5in]{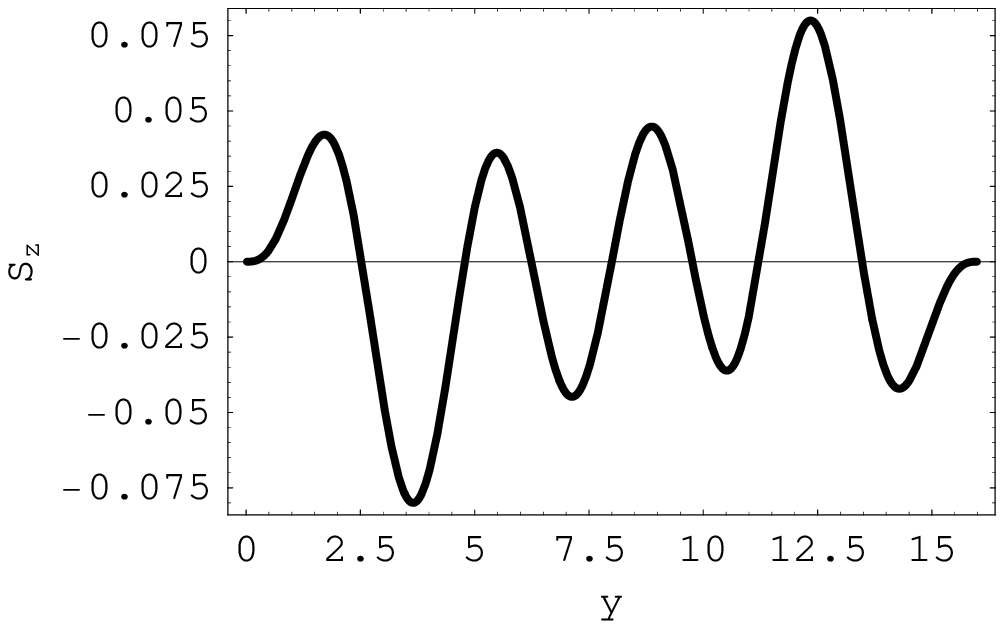}}
        \subfigure[]{
        \includegraphics[height=0.9in,width=1.5in]{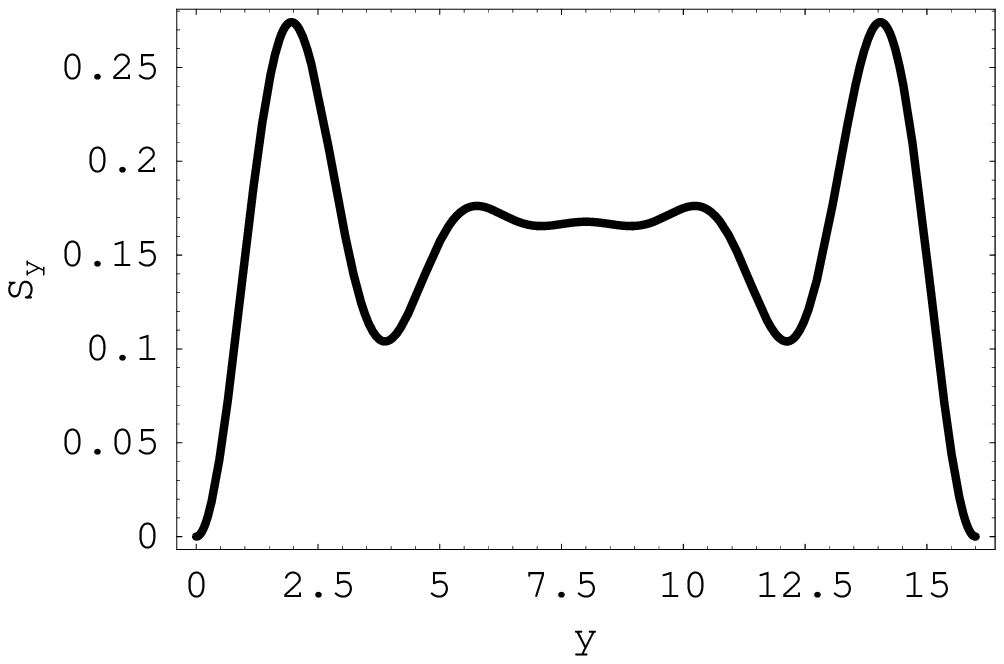}}
        \vspace{0.03in}
    \caption{Total local $S_zL/I$ and $S_yL/I$ as functions of y for $L=16/k_F$.
    In (a), $\lambda=0.01$. In (b), $\lambda=0.05$. In (c), $\lambda=0.1$.}
    \label{fig:totalL16}
\end{figure}
We have also obtained the local spin polarization for the case of
$L=16/k_F$. Figure~\ref{fig:totalL16} shows the results for
$\lambda=$0.01, 0.05, and 0.1. There are 10 eigenvalues for $k_x$
for each $\lambda$. Similarly to Fig.~\ref{fig:totalL8}, the
magnitude of spin polarization $S_zL/I$ near the edges increase as
$\lambda$ increases when $\lambda$ is small. However, when
$\lambda$ is large ($\lambda=0.1$), magnitude of the polarization $S_z$ near
the edges becomes smaller and large oscillations appears deep inside of the
sample. $S_y$ also increases as $\lambda$ increases. When
$\lambda=0.1$, in-plane spin polarization $S_y$ dominates
out-of-plane polarization $S_z$. Comparing with the results of
$L=8/k_F$, $S_zL/I$ is larger near the edges for $\lambda=$0.01 and 0.05. But when $\lambda=$0.1, it is smaller. $S_yL/I$ is larger for all $\lambda$'s we chose. We conclude that for
fixed finite sample size, the out-of-plane spin polarization $S_z$
dominates at relatively weak Rashba coupling and in-plane spin
polarization overwhelms for relatively strong Rashba coupling.
\begin{figure}[tbp]
\vspace{0.1in}
    \centering
    \subfigure[]{
        \includegraphics[height=1.5in,width=2.5in]{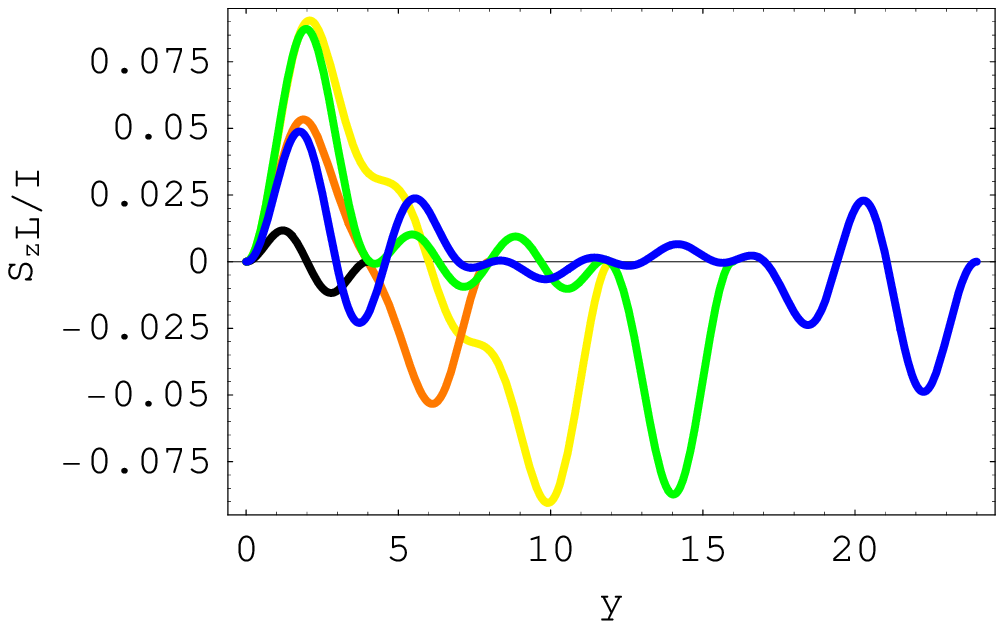}}
    \subfigure[]{
    \includegraphics[height=1.5in,width=2.5in]{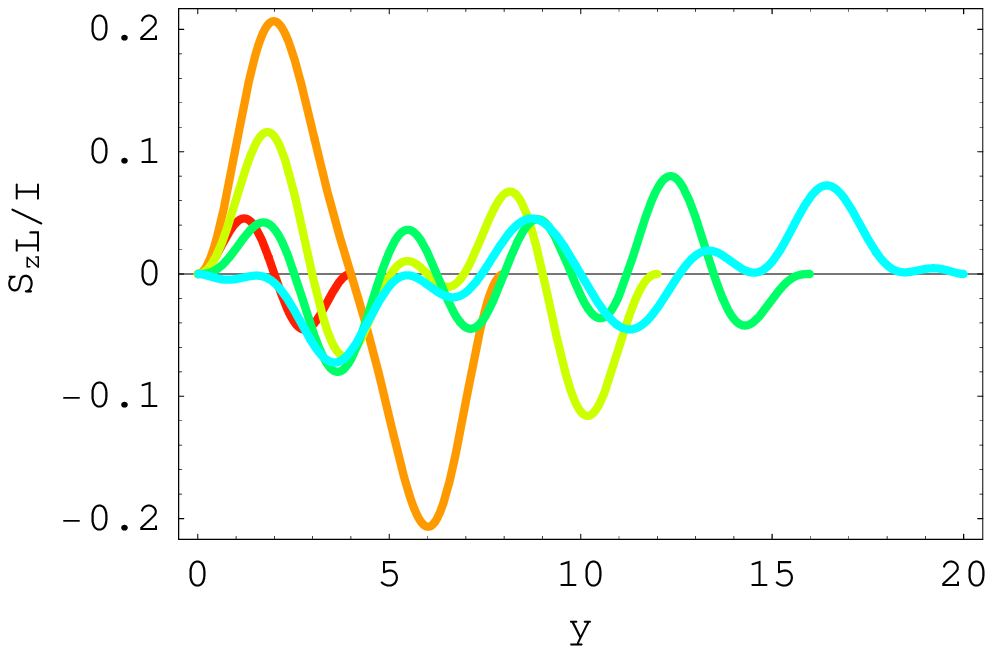}}
    \caption{Plots of $S_z(y)L/I$ for various $L$ with $\lambda=0.05$ [panel (a)] and $0.1$ [panel (b)].
    The value of $L$ of each curve can be identified from the right end of the curve. In panel (a), $L$=4,8,12,16,24; in panel (b), $L$=4,8,12,16,20. }
    \label{fig:zl}
\end{figure}
It is interesting to examine how the spin polarization changes as the
sample width $L$ increases. In Fig.~\ref{fig:zl}(a), we show
$S_zL/I$ for various $L$ with
$\lambda=0.05$. The peak magnitude of  $S_zL/I$ near the y=0
edge increases as $L$ increases at small $L$ and decreases as $L$ increases at large $L$. The plots for $\lambda=0.1$ in Fig.~\ref{fig:zl}(b) show a similar pattern, but the width, which has the biggest $S_zL/I$ near the edge, is shorter than that of $\lambda=0.05$.

In summary, we showed that in a 2-D narrow semiconductor strip with weak Rashba spin-orbit coupling,
local spin polarization could be induced by a steady longitudinal current, and it is originated from 
 the wave functions of the electrons at the Fermi level. The charge current along the
strip induces both out of plane and in plane local spin polarizations. Near the two edges, the  spin polarization $S_z$ has opposite sign, whereas $S_y$ has
the same sign. When the sample width $L$ increases, the peak magnitude of $S_zL/I$ near the edges increases at small $L$ and decreases at large $L$ for weak $\lambda$.  
And at large $L$, our numerical results indicate that $S_yL/I$
becomes dominant. From our scaling analysis based on varying $L$, the out-of-plane spin polarization is
important mainly in systems of mesoscopic sizes, and thus appears not to be associated with the SHE in bulk samples. 

Acknowledgement-We wish to thank J. Sinova for helpful discussion. This work is support ed by a grant from the Robert A. Welch Foundation and the Texas Center for Superconductivity at the University of Houston.

\end{document}